\documentclass[12pt]{article}
\usepackage[dvips]{graphicx}

\newcommand{\y}{y_i}
\newcommand{\phxx}{\mbox{\boldmath$\phi$}(x')}
\newcommand{\phx}{\mbox{\boldmath$\phi$}(x)}
\newcommand{\tphxx}{\mbox{\boldmath$\tilde\phi$}(x')}
\newcommand{\tphx}{\mbox{\boldmath$\tilde\phi$}(x)}
\newcommand{\phxi}{\mbox{\boldmath$\phi$}(x_i)}
\newcommand{\tphxi}{\mbox{\boldmath$\tilde\phi$}(x_i)}
\newcommand{\Kxx}{K(x,x')}

\newcommand{\Kinv}{K^{-1}\!(x,x')}
\newcommand{\w}{{\mathbf w}}
\newcommand{\tw}{{\mathbf {\tilde w}}}
\newcommand{\nsqw}{||\w||^2}
\newcommand{\xix}{\xi_i}
\newcommand{\hinge}{l}
\newcommand{\X}{X}

\newcommand{\heav}{H}
\newcommand{\ax}{\theta(x)}
\newcommand{\axi}{\theta(x_i)}
\newcommand{\ai}{\theta_i}
\newcommand{\axx}{\theta(x')}
\renewcommand{\a}{\mbox{$\theta$}}
\newcommand{\atr}{\mbox{\boldmath$\theta$}}
\newcommand{\atrt}{\mbox{\boldmath$\tilde\theta$}}
\newcommand{\att}{\tilde\theta}

\newcommand{\asvm}{\a^*}
\newcommand{\asvmx}{\theta^*\!(x)}
\newcommand{\asvmxi}{\theta^*_i}
\newcommand{\asvmxxi}{\theta^*(x_i)}

\newcommand{\norm}{\kappa}

\newcommand{\KK}{{\mathbf K}}
\newcommand{\MM}{{\mathbf M}_{\rm SV}}

\newcommand{\tr}{{\rm tr\,}}
\newcommand{\mident}{{\mathbf I}}
\newcommand{\cd}{\cdot}

\newcommand{\half}{{1\over 2}}

\newcommand{\lav}{\left\langle}
\newcommand{\rav}{\right\rangle}

\newcommand{\bi}{\begin{itemize}\setlength{\itemsep}{0pt}}
\newcommand{\ei}{\end{itemize}}
\newcommand{\be}{\begin{equation}}
\newcommand{\ee}{\end{equation}}
\newcommand{\bea}{\begin{eqnarray}}
\newcommand{\eea}{\end{eqnarray}}
\newcommand{\beastar}{\begin{eqnarray*}}
\newcommand{\eeastar}{\end{eqnarray*}}
\newcommand{\eq}[1]{~(\ref{#1})}

\def\(#1){(\ref{#1})}

\newcommand{\ie}{{\it i.e.}}
\newcommand{\eg}{{\it e.g.}}

\newcommand{\koff}{k_{\rm off}}
\newcommand{\eps}{\epsilon}
\newcommand{\eloo}{\eps_{\rm loo}}
\newcommand{\espan}{\eps_{\rm span}}
\newcommand{\egacv}{\eps_{\rm gacv}}
\newcommand{\KSV}{{\mathbf K}_{\rm SV}}
\newcommand{\T}{^{\rm T}}
\newcommand{\pv}{{\mathbf p}}

\begin{document}

\title{Model Selection for Support Vector Machine Classification}

\author{
Carl Gold\\
Computation and Neural Systems\\ 
California Institute of Technology, 139-74\\
Pasadena, CA 91125\\
Email {\tt carlg@caltech.edu}\\
\and
Peter Sollich\\
Department of Mathematics, King's College London\\
Strand, London WC2R 2LS, U.K.\\
Email {\tt peter.sollich@kcl.ac.uk}
}

\maketitle

\begin{abstract}
We address the problem of model selection for Support Vector Machine
(SVM) classification. For fixed functional form of the kernel, model
selection amounts to tuning kernel parameters and the slack penalty
coefficient $C$. We begin by reviewing a recently developed probabilistic
framework for SVM classification. An extension to the case of SVMs
with quadratic slack penalties is given and a simple approximation for
the evidence is derived, which can be used as a criterion for model
selection. We also derive the exact gradients of the evidence in terms
of posterior averages and describe how they can be estimated
numerically using Hybrid Monte Carlo techniques. Though
computationally demanding, the resulting gradient ascent algorithm is
a useful baseline tool for probabilistic SVM model selection, since it
can locate maxima of the exact (unapproximated) evidence. We then
perform extensive experiments on several benchmark data sets. The aim
of these experiments is to compare the performance of probabilistic
model selection criteria with alternatives based on estimates of the
test error, namely the so-called ``span estimate'' and Wahba's
Generalized Approximate Cross-Validation (GACV) error. We find that
all the ``simple'' model criteria (Laplace evidence approximations,
and the Span and GACV error estimates) exhibit multiple local optima
with respect to the hyperparameters. While some of these give
performance that is competitive with results from other approaches in
the literature, a significant fraction lead to rather higher test
errors. The results for the evidence gradient ascent method show that
also the exact evidence exhibits local optima, but these give test
errors which are much less variable and also consistently lower than
for the simpler model selection criteria.
\end{abstract}


\section{Introduction}

Support Vector Machines (SVMs) have emerged in recent years as
powerful techniques both for regression and classification. One of the
central open questions is model selection: how does one tune the
parameters of the SVM algorithm to achieve optimal generalization
performance? We focus on the case of SVM classification, where these
``hyperparameters'' include any parameters appearing in the SVM
kernel, as well as the penalty parameter $C$ for violations of the
margin constraint.

Our aim in this paper is twofold. First, we extend our work on
probabilistic methods for SVMs to the case of quadratic slack
penalties; we also develop a ``baseline'' algorithm which can be used
to find in principle exact maxima of the evidence. Second, we perform
numerical experiments on a selection benchmark data sets to compare
the model selection criteria derived from the probabilistic view of
SVMs with alternatives that directly try to optimize estimates of test
error. Our focus in these experiments is less on computational
efficiency, but rather on the relative merits of the methods in terms
of the resulting generalization performance.

We begin in Sec.~\ref{sec:svm} with a brief review of SVM
classification and of its probabilistic interpretation; the setup will
be such that the extension of the probabilistic point of view to the
quadratic penalty case requires only small changes compared to linear
penalty SVMs. In Sec.~\ref{sec:criteria} we review some criteria for
model selection that have been proposed based on approximations to the
test error. We also describe previous approximations to the evidence
for the linear penalty SVM, and then give an analogue for
quadratic penalty SVMs. Exact expressions for gradients of the
evidence with respect to the hyperparameters are then derived in terms
of averages over the posterior. Sec.~\ref{sec:methods} has a
description of the methods we use in our numerical experiments on
model selection, including the Hybrid Monte Carlo algorithm which we
use to calculate evidence gradients numerically. The results of our
experiments on benchmark data sets are discussed in
Sec.~\ref{sec:results}; we conclude in Sec.~\ref{sec:conclusion} with
a summary and an outlook towards future work.

\section{SVM classification}
\label{sec:svm}

In this section, we give a very brief review of SVM classification;
for details the reader is referred to recent textbooks or review
articles such as~\cite{Burges98,CriSha00}. We also sketch the
probabilistic interpretation of SVMs, from which we later obtain
Bayesian criteria for SVM model selection.

Suppose we are given a set $D$ of $n$ training examples $(x_i,\y)$
with binary outputs $\y=\pm 1$ corresponding to the two classes. The
basic SVM idea is to map the inputs $x$ to vectors $\phx$ in some
high-dimensional feature space; ideally, in this feature space, the
problem should be linearly separable. Suppose first that this is
true. Among all decision hyperplanes $\w\cd\phx+b=0$ which separate
the training examples (i.e.\ which obey $\y(\w\cd\phxi+b)>0$ for all
$x_i\in \X$, $\X$ being the set of training inputs), the SVM solution
is chosen as the one with the largest {\em margin}, i.e.\ the largest
minimal distance from any of the training examples. Equivalently, one
specifies the margin to be equal to 1 and minimizes the squared length
of the weight vector $\nsqw$~\cite{CriSha00}, subject to the
constraint that $\y(\w\cd\phxi+b)\geq 1$ for all $i$. The quantities
$\y(\w\cd\phxi+b)$ are again called margins, although for an
unnormalized weight vector they no longer represent geometrical
distances~\cite{CriSha00}. This leads to the following optimization
problem: Find a weight vector $\w$ and an offset $b$ such that
$\half\nsqw$ is minimized, subject to the constraint that
$\y(\w\cd\phxi+b)\geq 1$ for all training examples.

If the problem is not linearly separable, or if one wants to avoid
fitting noise in the training data, `slack variables' $\xix\geq 0$ are
introduced which measure how much the margin constraints are violated;
one thus writes $\y(\w\cd\phxi+b)\geq 1-\xix$.  To control the amount
of slack allowed, a penalty term $(C/p)\sum_i \xix^p$ is then added to the
objective function $\half\nsqw$, with a penalty coefficient
$C$. Common values for the exponent parameter are $p=1$ and $p=2$,
giving linear and quadratic slack penalties respectively. Training
examples with $\y(\w\cd\phxi+b)\geq 1$ (and hence $\xix=0$) incur no
penalty; the others contribute $(C/p)[1-\y(\w\cd\phxi+b)]^p$ each. This
gives the SVM optimization problem: Find $\w$ and $b$ to minimize
\be
\half\nsqw + C \sum_i \hinge_p(\y[\w\cd\phxi+b])
\label{basic_opt}
\ee
where $\hinge_p(z)$ is the loss function
\be
\hinge_p(z)=\frac{1}{p}(1-z)^p\heav(1-z)
\label{hinge}
\ee
The Heaviside step function $\heav(1-z)$ (defined as $H(a)=1$ for
$a\geq 0$ and $H(a)=0$ otherwise) ensures that this is zero for $z>
1$. For $p=1$, $\hinge_p(z)$ is called (shifted) hinge loss or soft
margin loss.

In the following we modify the basic SVM problem by adding the
quadratic term $\half b^2/B^2$ to\eq{basic_opt}, thus introducing a
penalty for large offsets $b$. A discussion of why this is reasonable,
certainly within a probabilistic view, can be found
in~\cite{Sollich02}; at any rate the standard formulation can always
be retrieved by making the constant $B$ large. We can now define an
augmented weight vector $\tw=(b/B,\w)$ and augmented feature space
vectors $\tphx = (B,\phx)$ so that the modified SVM problem is to find
a $\tw$ which minimizes
\be
\half||\tw||^2 + C \sum_i \hinge_p(\y\tw\cd\tphxi) 
\label{opt_no_b}
\ee
This statement of the problem is useful for the probabilistic
interpretation of SVM classification, of which more shortly. For a
practical solution, one uses Lagrange multipliers $\alpha_i$ conjugate
to the constraints $\y\tw\cd\tphxi\geq 1-\xix$ and finds in the
standard way (see e.g.~\cite{CriSha00}) that the optimal (augmented)
weight vector is $\tw^*=\sum_i \y\alpha_i\phxi$. For the linear
penalty case $p=1$, the $\alpha_i$ are found from
\be
\max_{0\leq \alpha_i\leq C}
\left(\sum_i \alpha_i - \half\sum_{i,j} \alpha_i\alpha_j\y y_j K_{ij}\right)
\label{linear_problem}
\ee
Here $K_{ij}=K(x_i,x_j)$ are the elements of the Gram matrix $\KK$,
obtained by evaluating
the {\em kernel} $K(x,x')=\tphx\cdot\tphxx = \phx\cdot\phxx+B^2$ for
all pairs of training inputs. The corresponding optimal latent or
discrimination function is $\asvmx=\tw^*\cdot \tphx = \sum_i
\y\alpha_i K(x,x_i)$. Only the $x_i$ with $\alpha_i>0$ contribute to
this sum; these are called support vectors (SVs). SVs fall into two
groups: If $\alpha_i<C$, one has $\y\asvmxi \equiv \y\asvmxxi = 1$;
we will call these the ``marginal SVs'' because their margins are
exactly at the allowed limit where no slack penalty is yet
incurred. For $\alpha_i=C$, on the other hand, $\y\asvmxi \leq 1$, and
these ``hard SVs'' are the points at which the slack penalty is
active. Non-SVs have large margins, $\y\asvmxi\geq 1$.

For the quadratic penalty case $p=2$, the $\alpha_i$ are obtained as
the solution of (see e.g.~\cite{CriSha00})
\be
\max_{0\leq \alpha_i}
\left(\sum_i \alpha_i - \half\sum_{i,j} \alpha_i\alpha_j\y y_j K^C_{ij}\right)
\label{quadratic_problem}
\ee
where $K^C_{ij}=K_{ij}+C^{-1}\delta_{ij}$. Apart from the replacement
of $K$ by $K^C$, this maximization problem is the same
as\eq{linear_problem} for the linear penalty case in the limit
$C\to\infty$ where no violations of the margin constraints are
allowed. There is now only one kind of SV, identified by
$\alpha_i>0$. It follows by differentiating\eq{quadratic_problem} that
for a SV one has $\y\sum_{j}\alpha_j y_j K^C_{ij} = 1$. The margin for
a SV is thus $\y\asvmxi = \y\sum_j \alpha_j y_j K_{ij} =
1-\alpha_i/C$, so that all SVs incur a nonzero slack penalty. Non-SVs
again have $\y\asvmxi\geq 1$.

We now turn to the probabilistic interpretation of SVM classification
(see Refs.~\cite{Sollich02,Sollich99_ICANN_SVM,Sollich00_NIPS_SVM} and
the works quoted below). The aim of such an interpretation is to allow
the application of Bayesian methods to SVMs, without modifying the
basic SVM algorithm which already has a large user community. (An
alternative philosophy would be to consider similar inference
algorithms which share some of the benefits of SVMs but are
constructed directly from probabilistic models; Tipping's Relevance
Vector Machine~\cite{Tipping00} is a successful example of this.) One
regards\eq{opt_no_b} as defining a negative log-posterior probability
for the parameters $\tw$ of the SVM, given a training set $D$. The
conventional SVM classifier is then interpreted as the maximum a
posteriori (MAP) solution of the corresponding probabilistic inference
problem. The first term in\eq{opt_no_b} gives the prior
$Q(\tw)\propto\exp(-\half||\tw||^2)$. This is a Gaussian prior on
$\tw$; the components of $\tw$ are uncorrelated with each other and
have unit variance. Because only the `latent function' values
$\ax=\tw\cd\tphx$---rather than $\tw$ itself---appear in the second,
data dependent term of\eq{opt_no_b}, it makes sense to express the
prior directly as a distribution over these. The $\ax$ have a joint
Gaussian distribution because the components of $\tw$ do, with
covariances given by
\[
\lav\ax\axx\rav = \lav(\tphx\cd\tw)(\tw\cd\tphxx)\rav = K(x,x')
\]
The SVM prior is therefore simply a {\em Gaussian process} (GP) over
the functions $\a$, with zero mean and with the kernel $K(x,x')$ as
covariance function. This link between SVMs and GPs has been pointed
out by a number of authors, \eg~\cite{Seeger00,OppWin99,OppWin00}. It
can be understood from the common link to reproducing kernel Hilbert
spaces~\cite{Wahba97}, and can be extended from SVMs to more general
kernel methods~\cite{JaaHau98}. For connections to regularization
operators see also~\cite{SmoSchMul98}. A nice introduction to
inference with Gaussian processes can be found in
Ref.~\cite{Williams98}.

The second term in\eq{opt_no_b} similarly becomes a (negative)
log-likelihood if we define the probability of obtaining output $y$
for a given $x$ (and $\a$) as
\be
Q(y\!=\!\pm 1|x,\a) = \norm(C) \exp[-C\hinge_p(y\ax)]
\label{lh}
\ee
The constant factor $\norm(C)$ is determined from
$\norm^{-1}(C)=\max_z
\{e^{-C\hinge_p(z)}+e^{-C\hinge_p(-z)}\}$ to ensure that
$\sum_{y=\pm 1} Q(y|x,\a)\leq 1$. In the linear penalty case this
gives $\norm(C)=1/[1+\exp(-2C)]$; for the quadratic penalty SVM, the
maximum in the definition of $\norm^{-1}(C)$ is assumed at a value of
$\theta$ obeying $\theta=\tanh(C\theta)$ and so $\norm(C)$ can easily
be found numerically. The likelihood for the complete data set (more
precisely, for the training outputs $Y=(y_1\ldots y_n)$ given the
training inputs $X$) is then
\[
Q(Y|X,\a)=\prod_i Q(y_i|x_i,\a)
\]
With these definitions eq.\eq{opt_no_b} is, up to unimportant
constants, equal to the log-posterior\footnote{%
In~(\protect\ref{log_post}) the unrestricted sum over $x$ runs over
all possible inputs, and $\Kinv$ are the elements of the inverse of
$\Kxx$, viewed as a matrix. We assume here that the input domain is
discrete. This avoids mathematical subtleties with the definition of
determinants and inverses of operators (rather than matrices), while
maintaining a scenario that is sufficiently general for all practical
purposes.
}
\be
\ln Q(\a|X,Y) = -\half\sum_{x,x'} \ax \Kinv \,\axx -
C\sum_i \hinge_p(\y\axi) + \mbox{const}
\label{log_post}
\ee
By construction, the maximum of $\asvmx$ gives the conventional SVM
classifier, and this is easily verified explicitly~\cite{Sollich02}.

The probabilistic model defined above is not normalized, since
$\sum_{y=\pm 1}Q(y|x,\a)<1$ for generic values of $\ax$. The
implications of this have been previously discussed in
detail~\cite{Sollich02}. The normalization of the model is important
for the theoretical justification of tuning hyperparameters via
maximization of the data likelihood or ``evidence''. Nevertheless,
experiments in~\cite{Sollich02} showed that promising results for
hyperparameter optimization could be obtained also with the
unnormalized version of the model. We will therefore, in common with
other work on probabilistic interpretations of
SVMs~\cite{Seeger00,OppWin99,OppWin00,Kwok99,Kwok00}, disregard the
normalization issue 
from now on. We will also focus on SVM classifiers constructed
from radial basis function (RBF) kernels
\be
K(x,x')=k_0 \exp\left[-\sum_a \frac{(x^a-x'^a)^2}{2
l_a^2}\right]+\koff
\label{rbf}
\ee
where the $x^a$ are the
different input components, $k_0$ is the kernel amplitude and $\koff$
the kernel offset; $\koff$ corresponds to the term $B^2$ discussed
above that arises by incorporating the offset $b$ into the
kernel. Each input dimension has associated with it a length scale
$l_a$. Since in the probabilistic interpretation $K(x,x')$ is the
prior covariance function of the latent function $\ax$, each $l_a$
determines the distance in the $x^a$-direction over which $\ax$ is
approximately constant; large $l_a$ correspond to an input component
of little relevance (see e.g.~\cite{Neal96}).

\section{Model selection criteria}
\label{sec:criteria}

\subsection{Error bounds and approximations}
\label{errbounds}

Model selection aims to tune the hyperparameters of SVM classification
(the penalty parameter $C$ and any kernel parameters) in order to
achieve the lowest test error $\eps$, \ie\ the lowest probability of
misclassification of unseen test examples. The test error is not
observable directly, and so one is lead to use bounds or
approximations as model selection criteria. The simplest such
bounds~\cite{Vapnik95,Vapnik98} which have been applied as model
selection criteria~\cite{CriCamSha99,ChaVap00} are expressed in
terms of the quantity
\be
\frac{R^2}{n}\sum_i \alpha_i
\label{margin_bound}
\ee
Here $R$ is the radius of the smallest ball in feature space
containing all training examples, while $\sum_i \alpha_i$ can be shown
to equal the inverse square of the distance between the separating
hyperplane and the closest training points. For RBF kernels, $R$ is
bounded by a constant since every input point has the same squared
distance $\tphx\cdot\tphx = K(x,x)$ from the origin.

More recent work has shown that better bounds and approximations can
be obtained for the leave-out-out error $\eloo$. If $\a^i(x)$ is the
latent function obtained by training the SVM classifier on the data
set with example $(x_i,\y)$ left out, then $\eloo$ is the probability
of misclassification of the left-out example if this procedure is
applied to each data point in turn,
\be
\eloo=\frac{1}{n}\sum_i H(-\y\a^i_i)
\label{eloo_def}
\ee
where we have abbreviated $\a^i_i\equiv \a^i(x_i)$.  Averaged over
data sets this is an unbiased estimate of the average test error that
is obtained from training sets of $n-1$ examples. This says nothing
about the variance of this estimate; nevertheless, one may hope that
$\eloo$ is a reasonable proxy for the test error that one wishes to
optimize. (This is in contrast to the training error, \ie\ the
fraction of all $n$ training examples misclassified when training on
the complete data set, which is general a strongly biased estimate of
test error.) For large data sets, $\eloo$ is time-consuming to compute
and one is driven to look for cheaper bounds or approximations. Since
removing non-SVs from the data set does not change the SVM classifier,
a trivial bound on $\eloo$ is the sum of the training error and the
fraction of support vectors, both obtained when training on all $n$
examples. To get better bounds, one writes
\[
\eloo = \frac{1}{n}\sum_i H\left(\y[\asvmxi-\a^i_i] - \y\asvmxi\right)
\]
which shows that an upper bound on $\y[\asvmxi-\a^i_i]$ will give an
upper bound on $\eloo$.  Jaakkola and Haussler proved a bound of this
form, $\y[\asvmxi-\a^i_i]\leq \alpha_i K_{ii}$; as before, the
$\alpha_i$ are those obtained from training on the full data set.
More sophisticated bounds were given by Chappelle and
Vapnik~\cite{VapCha00,ChaVap00,ChaVapBouMuk02} in terms of what
they called the ``span''. We focus on the case of quadratic penalty
SVMs, where the span estimates are simplest to state. In the simplified
version of Ref.~\cite{ChaVapBouMuk02}, and adapting to our formulation with
the offset $b$ incorporated into the kernel, the span $S_i$ for a
support vector can be defined as
\be
S^2_i = \min_\lambda \sum_{j,k} \lambda_j \lambda_k K^C_{jk}
\label{span_def}
\ee
where the minimum is over all $\lambda=(\lambda_1\ldots\lambda_n)$
with $\lambda_i=-1$, and $\lambda_j=0$ whenever $\alpha_j=0$. With
this definition, one can calculate $\y[\asvmxi-\a^i_i]$ exactly
under the assumption that dropping the point $x_i$ from the training
set leaves the ``SV set'' unchanged, in the sense that no new SVs
arise in the new classifier and that all old SVs remain. One thus
finds $\y[\asvmxi-\a^i_i]=\alpha_i(S^2_i-1/C)$. $S^2_i$ can
also be worked out explicitly as
$S^2_i=1/[(\KSV+\mident/C)^{-1}]_{ii}$, where $\KSV$ is the Gram matrix
$\KK$ restricted to the SVs and $\mident$ is the unit matrix. (The
same result was obtained by Opper and Winther~\cite{OppWin00} using a
slightly different approach.)  Using finally that
$\y\asvmxi=1-\alpha_i/C$ for quadratic penalty SVMs,
one thus has
\be
\eloo\approx \espan
= \frac{1}{n}\sum_i H\left(\alpha_i S_i^2 - 1\right)
, \qquad S_i^2 = 1/[(\KSV+\mident/C)^{-1}]_{ii}
\label{span}
\ee
This is only an approximation because the assumption of an unchanged SV
set will not hold for every SV removed from the training set.

The span estimate\eq{span} of leave-one-out error has the undesirable
property of being discontinuous as hyperparameters are varied, making
numerical optimization difficult. The discontinuity arises from the
discontinuity in the Heaviside step function $H$, and from the fact
that the size of the matrix $\KSV$ changes as training examples
enter or leave the set of SVs. To get around this~\cite{ChaVapBouMuk02},
one can approximate $H(z)$ by a sigmoidal function
$1/[1+\exp(-c_1x+c_2)]$ and smooth the span by adding a penalty that
forces any nonzero $\lambda_j$ to go to zero when $\alpha_j\to
0$. This gives the modified span definition
\[
S^2_i = \min_\lambda \sum_{j,k} \lambda_j \lambda_k K^C_{jk} + \eta
\sum_{j\neq i} \frac{\lambda_j^2}{\alpha_j}
\]
with the minimum taken over the same $\lambda$ as
in\eq{span_def}. Explicitly, one finds
\[
S^2_i = \frac{1}{[(\KSV+\mident/C+
\eta{\mathbf A}_{\rm SV}^{-1})^{-1}]_{ii}} -
\frac{\eta}{\alpha_i}
\]
where ${\mathbf A}_{\rm SV}$ is the diagonal matrix containing the
nonzero $\alpha_i$. This is easily seen to be continuous even when the
set of SVs changes as hyperparameters are varied. For $\eta\to 0$ one
recovers the original span definition\eq{span_def}; for
$\eta\to\infty$, on the other hand, $S^2_i\to K^C_{ii}=K_{ii}+1/C$ and one
recovers the Jaakkola and Haussler bound. Overall, the smoothed span
estimate for $\eloo$ contains three smoothing parameters $c_1$, $c_2$
and $\eta$.

For linear penalty SVMs, Wahba~\cite{Wahba97} considered a modified
version of $\eloo$, obtained by replacing the Heaviside step function
$H(-z)$ in\eq{eloo_def} by the hinge loss $\hinge_1(z)=(1-z)H(1-z)$; since
$\hinge_1(z)\geq H(-z)$, this actually gives an upper bound on
$\eloo$. Wahba's GACV (generalized approximate cross-validation)
estimate for this modified $\eloo$ is
\be
\egacv = \frac{1}{n}\sum_i \left[l_1(\y\asvmxi) + \alpha_i K_{ii}
f(\y\asvmxi)\right]
\label{eq:gacv}
\ee
where
\[
f(z) =\left\{
\begin{array}{ll}
2\qquad & x < -1 \\
1\qquad & -1 \leq x \leq 1 \\
0\qquad & x>1
\end{array}
\right.
\]
The first term in the sum in\eq{eq:gacv} would just give the naive
estimate of the (modified) $\eloo$ from the performance on the
training set; the second term effectively corrects for the bias in
this estimate. Because of the nature of the function $f$, $\egacv$ can
exhibit discontinuities as hyperparameters are varied and this has to
be taken into account when minimizing it numerically.

\subsection{Approximations to the evidence}
\label{approxevid}

From a probabilistic point of view, it is natural to tune
hyperparameters to maximize the likelihood of the data $Q(Y|X)$. By
definition, $Q(Y|X)=\int d\a\, Q(Y|X,\a)Q(\a)$ where the integration is
over the values $\ax$ of the latent function $\a$ at all different
input points $x$. The likelihood $Q(Y|X,\a)$ only depends on the
values $\ai\equiv\a(x_i)$ of $\a$ at the training inputs; all other
$\ax$ can be integrated out trivially, so that
\[
Q(Y|X) = \int d\atr\, Q(Y|X,\atr) Q(\atr)
\]
where now the integral is over the $n$-dimensional vector
$\atr=(\a_1\ldots \a_n)$. Because $Q(\a)$ is a zero mean Gaussian
process, the marginal $Q(\atr)$ is a zero mean Gaussian distribution
with covariance matrix $\KK$. The evidence is therefore
\be
Q(Y|X) = |2\pi\KK|^{-1/2}\norm^n(C) \int d\atr\, 
\exp\left[-\half\atr\T\KK^{-1} \atr - \sum_i C\hinge_p(\y\ai)\right]
\label{eq:pyx1}
\ee
This n-dimensional integral is in general impossible to carry out
exactly. But it can be approximated by expanding the exponent around
its maximum $\atr^{\ast}$, the solution of the SVM optimization
problem. For the linear penalty case, this requires some care: because
of the kink in the hinge loss $\hinge_1(z)$, the $\ai$ corresponding
to marginal SVs have to be treated separately. The result for the
normalized log-evidence, suitably smoothed to avoid spurious
singularities, is~\cite{Sollich02}
\begin{equation}\label{eq:lapnv}
E(Y|X) \equiv \frac{1}{n}\ln Q(Y|X) = \frac{1}{n}\ln{Q^{\ast}(Y|X)} - \frac{1}{2n}\ln{\det({\mathbf{I}+\mathbf{L}_m\mathbf{K}_m})}
\end{equation}
where $\mathbf{K}_m$ is the sub-matrix of the Gram matrix
corresponding to the marginal SVs, $\mathbf{L}_m$ is the diagonal
matrix with entries $2\pi[\alpha_i(C-\alpha_i)/C]^2$ and
\begin{equation}\label{p*yx}
  \frac{1}{n}\ln{Q^{\ast}(Y|X)}=-\frac{1}{2n}\sum_{i}\alpha_i y_i\theta^\ast_i
 - \frac{C}{n}\sum_i \hinge_p(y_i\theta^\ast_i)+\ln{\norm(C)}
\end{equation}
The approximation\eq{eq:lapnv} is computationally efficient because it
only involves 
calculating the determinant of a single matrix of size equal to the
number of marginal SVs. A related approximation was proposed by
Kwok~\cite{Kwok00}. He suggested to smooth the kink in the hinge loss
by using a sigmoidal approximation for the Heaviside function, giving
$l_1(z) \approx s(z)=(1-z)/[1+\exp[-c(1-z)]$ with a smoothing
parameter $c$. This has the disadvantage that the SVM solution $\asvm$
is no longer a maximum of the smoothed posterior. The analogous result
to\eq{eq:lapnv} also involves, instead of $\mathbf{K}_m$ and
$\mathbf{L}_m$, the whole Gram matrix and a diagonal matrix with
entries $Cs''(\y\asvmxi)$, respectively. One thus needs either to
evaluate a large determinant of size $n$, or---somewhat
arbitrarily---to truncate small values of $s''(\y\asvmxi)$ to zero to
reduce the size of the problem. We therefore do not consider this
approach further.

%
%

An approximation similar to\eq{eq:lapnv} can easily be derived for the
quadratic penalty case. The loss function $\hinge_2(z)$ now has a
continuous first derivative and so all $\ai$ can be treated on the
same footing.  The derivation of the Laplace approximation is thus
standard, and involves the Hessian of the log-likelihood at the
maximum $\atr^\ast$; the resulting approximation to the (normalized
log-) evidence is
\begin{equation}\label{eq:laplace_qu}
E(Y|X) = \frac{1}{n}\ln{Q^{\ast}(Y|X)} - \frac{1}{2n}\ln{\det({\mident+\MM\KSV})}
\end{equation}
where $\MM$ is a diagonal matrix containing the second derivatives
$C\hinge''_2(\y\asvmxi)$ of the loss function evaluated for all the
SVs. The calculation of $E(Y|X)$ according to\eq{eq:laplace_qu}
requires only the determinant of a matrix whose size is the number of
SVs, again expected to be manageably small. The matrix $\MM$ as
defined above is just a multiple of the unit matrix, since
$\hinge''_2(z)=H(1-z)$ and $z=\y\asvmxi<1$ for SVs. However, the step
discontinuity in $\hinge''_2(z)$ at $z=1$ has the undesirable
consequence that the Laplace approximation to the evidence will jump
discontinuously when one or several of the $\alpha_i$ reach zero as
hyperparameters are varied. We therefore smooth the result by
replacing $\hinge''_2(z)=H(1-z)$ in the definition of $\MM$ by the
approximation $\hinge''_2(z)\approx\exp[-a/(1-z)]$ for $z<1$ (and $0$
for $z\geq 1$). This is smooth at $z=1$ and also has continuous
derivatives of all orders at this point. The value of $a$ determines
the range of values of $z=\y\ai$ around 1 for which the smoothing is
significant, with $a\to 0$ recovering the Heaviside step function.

\subsection{Evidence gradients}

Beyond the relatively simple approximations to the evidence derived
above, it is difficult to obtain accurate numerical estimates of the
evidence. This is a well-known general problem: while averages over
probability distributions are straightforward to obtain, normalization
constants for such distributions -- such as the evidence, which is the
normalization factor for the posterior -- require much greater
numerical effort (see \eg~\cite{Neal93}). To avoid this problem, one
can estimate the
gradients of the evidence with respect to the hyperparameters and use
these in a gradient ascent algorithm, without ever calculating the
value of the evidence itself. As we show in this section, these
gradients can be expressed as averages over the posterior
distribution, which one can then estimate by sampling as explained in
Sec.~\ref{hmc}.

Starting from eq.\eq{eq:pyx1} we can find the derivative of the
normalized log-evidence $E(Y|X)=n^{-1}\ln Q(Y|X)$ w.r.t.\ the penalty
(or noise) parameter C:
\begin{eqnarray}
\frac{\partial}{\partial{}C}E(Y|X)
&=&\frac{\partial\ln\kappa(C)}{\partial{}C}-
\frac{\int
d{\atr}\, Q(\atr) \sum_{i}\hinge_p(y_{i}\theta_i) \exp\left[-C\sum_{i}\hinge_p(y_{i}\theta_i)\right]}{\int
d{\atr}\, Q(\atr) \exp\left[-C\sum_{i}\hinge_p(y_{i}\theta_i)\right]} \nonumber \\
&=&\frac{\partial\ln\kappa(C)}{\partial{}C} -
\left \langle \frac{1}{n}\sum_{i}\hinge_p(y_{i}\theta_{i})
\right \rangle \label{eq:cgrad}
\end{eqnarray}
where the average is, as expected, over the posterior
$Q(\atr|D)\propto Q(Y|X,\atr)Q(\atr)$.
Similarly, the derivative of the log-evidence w.r.t.\ 
any parameter $\lambda$ appearing in the kernel is
\begin{eqnarray}
\frac{\partial}{\partial{}\lambda}E(Y|X)&=&
-\frac{1}{2n}\frac{\partial}{\partial{}\lambda}
\ln |2\pi\mathbf{K}| -
\frac{\int d\atr\,
\frac{1}{2}\atr\T\frac{\partial}{\partial{}\lambda}\mathbf{K}^{-1}\atr
\exp\left[-\half\atr\T\KK^{-1} \atr - \sum_i C\hinge_p(\y\ai)\right]}
{\int d\atr\, 
\exp\left[-\half\atr\T\KK^{-1} \atr - \sum_i C\hinge_p(\y\ai)\right]}
\nonumber\\
&=&
-\frac{1}{2n}\tr\left(
\frac{\partial\mathbf{K}}{\partial{}\lambda}\mathbf{K}^{-1}\right) +
\frac{1}{2n}\lav
\atr\T\mathbf{K}^{-1}\frac{\partial\mathbf{K}}{\partial{} \lambda}
\mathbf{K}^{-1}\atr \rav\nonumber
\\
&=&
-\frac{1}{2n}\tr\left(
\frac{\partial\mathbf{K}}{\partial{}\lambda}\mathbf{K}^{-1}\lav
\mathbf{I} - \atr\atr\T\mathbf{K}^{-1}\rav \right) \label{eq:ddl1}
\end{eqnarray}
Numerical evaluation of this expression as it stands would be unwise,
since the difference $\lav \mathbf{I} - \atr\atr\T\mathbf{K}^{-1}\rav$
can be much smaller than the two contributions individually; in fact,
for $n=0$ we know that it is exactly zero. It is better to
rewrite\eq{eq:ddl1}, using the fact that the elements of the matrix
$\mathbf{I} - \atr\atr\T\mathbf{K}^{-1}$ can be obtained as
\[
\delta_{ij} - \theta_i(\mathbf{K}^{-1}\atr)_j 
= \exp\left[\frac{1}{2}\atr\T\mathbf{K}^{-1}\atr\right]
\frac{\partial}{\partial\theta_j}\theta_i
\exp\left[-\frac{1}{2}\atr\T\mathbf{K}^{-1}\atr\right]
\]
The posterior average can thus be worked out using integration by
parts, giving
\[
\lav \delta_{ij} - \theta_i(\mathbf{K}^{-1}\atr)_j \rav
=
\lav C \hinge_p'(y_j\theta_j)y_j \theta_i \rav
\]
If we define the matrix $\mathbf{Y}$ as the diagonal matrix with
entries $y_i$ so that $(\mathbf{Y}\atr)_i = y_i \theta_i$, and denote
by $\hinge_p'(\mathbf{Y\atr})$ the vector with entries $\hinge_p'(y_i\theta_i)$,
then this can be written in the compact form
\[
\lav\mathbf{I} -
\atr\atr\T\mathbf{K}^{-1}\rav
= C\lav \atr [l'_p(\mathbf{Y}\atr)]\T\mathbf{Y}) \rav \label{eq:ddl2}
\]
Combining this with equation\eq{eq:ddl1}, one has finally
\be
\frac{\partial}{\partial{}\lambda}E(Y|X)
=
-\frac{C}{2n}\lav [\hinge'_p(\mathbf{Y}\atr)]\T \mathbf{Y}
\frac{\partial\mathbf{K}}{\partial{}\lambda}\mathbf{K}^{-1} \atr
\rav
\label{eq:lgrad}
\ee
This expression appears to require the inverse $\KK^{-1}$ of the Gram
matrix, which for large $n$ would be computationally expensive to
evaluate; however, as described in Sec.~\ref{hmc} the sampling from
the posterior using Hybrid Monte Carlo can be arranged so
that samples of both $\atr$ and $\KK^{-1}\atr$ are obtained without
requiring explicit matrix inversions.


\section{Numerical methods}
\label{sec:methods}

In Sec.~\ref{sec:results} below, we report results from SVM model
selection experiments using the criteria described
above. Specifically, for linear penalty SVMs we compare maximizing the
Laplace approximation to the evidence $E(Y|X)$, eq.\eq{eq:lapnv}, with
minimizing Wahba's $\egacv$, eq.\eq{eq:gacv}; for quadratic penalty
SVMs we again use the relevant approximation to the evidence,
eq.\eq{eq:laplace_qu}, and contrast with minimization of the span
estimate $\espan$ of the leave-one-out error, eq.\eq{span}. These
four model selection criteria are ``simple'' in the sense that they
can be evaluated explicitly at moderate computational cost. In order
to be able to compare the different criteria directly, and because one
of them ($\egacv$) has possible discontinuities as a function of the
hyperparameters, we use a simple greedy random walk algorithm for
optimization that is described in Sec.~\ref{sec:ztmc}. For the other
three criteria, more efficient gradient-based optimization algorithms
can be designed~\cite{ChaVapBouMuk02} but since our focus here is
not on computational efficiency we do not consider these.

For linear penalty SVMs we also studied evidence optimization using
numerical estimates of the evidence
gradients~(\ref{eq:cgrad},\ref{eq:lgrad}). The Monte Carlo method used
to perform the necessary averages over the posterior is outlined in
Sec.~\ref{hmc}, while Sec.~\ref{sec:gradient_ascent} describes the
details of the gradient ascent algorithm. Note that our use of
evidence gradients provides a baseline for model selection methods
based on approximations to the evidence since it locates, 
up to small statistical errors from the Monte
Carlo sampling of posterior averages, a local maximum of the exact
evidence.

In all experiments using approximations to the test error ($\espan$
and $\egacv$) as model selection criteria, the hyperparameters being
optimized were the parameters of the RBF kernel\eq{rbf}, \ie\ the
amplitudes $k_0$, $\koff$ and the logarithms of the length scales $l_a$
(one per input dimension). The penalty parameter $C$ can be fixed to
\eg\ $C=1$ since these criteria depend only on the properties of the
SVM solution (\ie\ the maximum of the posterior, rather than the whole
posterior distribution). This SVM solution only depends on the product
$C\Kxx$ rather than $C$ and the kernel individually, as one easily
sees from~(\ref{linear_problem},\ref{quadratic_problem}) or
equivalently from\eq{log_post}. In contrast, the evidence\eq{eq:pyx1}
takes into account both the position of the posterior maximum and the
shape of the posterior distribution around this maximum; the latter does
depend on $C$. We therefore include $C$ as a hyperparameter to be
optimized in evidence maximization. The SVM predictor of the final
selected model will of course again be dependent only on the product
$C\Kxx$; but the value of $C$ itself would be important \eg\ for the
determination of predictive class probabilities. This issue, which we
do not pursue here, is discussed in detail in Ref.~\cite{Sollich02}.

\subsection{Optimization of ``simple'' model selection criteria}
\label{sec:ztmc}

We optimized the simple model selection criteria using a simple greedy
random walk, or ``zero temperature Monte Carlo'' search.  This is a
simple adaptation of the common Metropolis Algorithm (see
e.g.~\cite{Krauth98}) used to sample from a probability distribution;
in the zero temperature limit the algorithm reduces to repeatedly
adding a small step (which we take to be Gaussian) to each parameter,
recalculating the quantity being optimized, and moving to the new
point if and only if the new point yields a better value (higher for
the evidence, and lower for error estimates). The randomness in the
algorithm may appear disadvantageous in terms of computational
efficiency; but for our purposes, it is actually helpful since it
allows us to assess whether the model selection criteria in question
have a number
of local optima or a single (global) optimum. It also made further
randomization over the initial hyperparameter values unnecessary, and
so the experiments with the simple model selection criteria were all
started with a fixed set of initial values for the SVM
hyperparameters.  A few preliminary trials were used to choose initial
values with an appropriate order of magnitude, and all results
reported were initialized with the hyperparameters $C=1$, $l_a = 1$
for all length scales, $k_0=1$ and $\koff=0.1$.

The span error estimate and the Laplace evidence for quadratic
loss each have additional smoothing parameters that had to be selected
($c_1$, $c_2$ and $\eta$ in the span estimate, $a$ for the Laplace
evidence; see Secs.~\ref{errbounds} and~\ref{approxevid}
respectively). Appropriate values for these parameters were found by
a simple (log) line search in the parameter values.  These tests were
not done extensively, but the results for SVM model selection did not
seem to depend strongly on the values of these parameters as long as
they were of a reasonable order of magnitude. For all
tests presented here the values used were $c_1=5$, $c_2=0$,
$\eta=1$ for the span
estimate, and $a=0.1$ for the Laplace evidence with quadratic loss.

In the greedy random walk algorithm, the step size used for each
hyperparameter was 
adapted separately by measuring the acceptance rate for proposed
changes in the parameter and scaling the step size up or down to keep
the acceptance rate close to 50\%.  Thus a decreasing step size can be
taken as one measure of how well the process has converged to an
optimum. The search is terminated when either the step size has become
very small, or the change to the criterion being optimized becomes
very small.  It was also found during experimentation that a useful
addition to the basic algorithm was to enforce minimum and maximum
values of the hyperparameters.  Without such bounds the algorithm
would occasionally get ``stuck'' in a plateau region of the model
selection criterion
where one or more hyperparameters were either very large or very
small. Note 
that for the kernel hyperparameters steps in the random walk were
taken in the natural logarithm of the hyperparameter values, as these
scale parameters were expected to show a significant range of variation.
Steps for $C$ were taken in a linear scale, reflecting the smaller
range of variation.

\subsection{Estimating evidence gradients}
\label{hmc}

We used Hybrid Monte Carlo (HMC, see \eg~\cite{Neal93}) to estimate
the posterior averages required in the expressions\eq{eq:cgrad}
and\eq{eq:lgrad} for the exact evidence gradients. The HMC algorithm
is a standard technique from statistical physics that works by
simulating a stochastic dynamics with a Hamiltonian ``energy'' defined
by the target distribution plus a ``momentum'', or kinetic energy
term. Denoting the momentum variables $\pv$, the Hamiltonian we choose
for our case is
\begin{equation}\label{eq:ham}
{\mathcal H}(\atr,\pv)= \frac{1}{2}\mathbf{p}^T \KK\mathbf{p} + \half\atr\T
\KK^{-1}\atr + V(\atr), \qquad
V(\atr) = C \sum_i \hinge_p(y_i\theta_i)
\end{equation}
and the corresponding ``Boltzmann'' distribution
$P(\atr,\pv)\propto\exp[-{\mathcal H}(\atr,\pv)]\propto \exp(\half \pv\T\KK\pv)
Q(\atr|D)$ factorizes over $\atr$ and $\pv$, so that samples from
$Q(\atr|D)$ can be obtained by sampling from $P(\atr,\pv)$ and
discarding the momenta $\pv$. The $\pv$ are nevertheless important for
the algorithm, since they help to ensure a representative sampling of
the posterior. An update step in the HMC algorithm consists of two
parts. First, one updates a randomly chosen momentum variable $p_i$ by
Gibbs sampling according to the Gaussian distribution
$\exp(-\frac{1}{2}\mathbf{p}\T \KK\mathbf{p})$; this will in general
change the value of the Hamiltonian. Second, one changes both $\atr$
and $\pv$ by 
moving along a Hamiltonian trajectory for some specified ``time''
$\tau$; the trajectory is determined by solving an appropriately
discretized version of the differential equations
\begin{eqnarray}
\frac{d\a_i}{d\tau} &=& \frac{\partial {\mathcal H}}{\partial p_i} = (\KK\pv)_i
\label{theta_i}\\
\frac{dp_i}{d\tau} &=& - \frac{\partial {\mathcal H}}{\partial \a_i} = -
(\KK^{-1}\atr)_i - \frac{\partial V(\atr)}{\partial \a_i}
\label{p_i}
\end{eqnarray}
For an exact solution of these equations, ${\mathcal H}$ would remain constant;
due to the discretization, small changes in ${\mathcal H}$ are possible and one
accepts the update of $\atr$ and $\pv$ from the beginning to the end
of the trajectory with the usual Metropolis acceptance rule. Iterating
these steps the algorithm will, after some initial equilibration
period, produce samples from $P(\atr,\pv)$.
%
%

The occurrence of $\KK^{-1}$ in\eq{p_i} is inconvenient. We circumvent
this by introducing $\atrt=\KK^{-1}\atr$; $\atr$ is initialized to
the SVM solution $\atr^*$, since then the corresponding $\atrt$ is obtained
trivially as $\att_i=y_i\alpha_i$ without requiring matrix
inversions. The Hamiltonian equations~(\ref{theta_i},\ref{p_i}) simplify to
\begin{eqnarray}
\frac{d\att_i}{d\tau} &=& p_i
\nonumber\\
\frac{dp_i}{d\tau} &=& - \att_i - \frac{\partial V(\atr)}{\partial \a_i}
\nonumber
\end{eqnarray}
and the simple form of the first equation is in fact what motivated
our choice of the momentum-dependent part of $H$, eq.\eq{eq:ham}. The
correspondence between $\atrt$ and $\atr$ is maintained by updating
$\atr=\KK\atrt$ whenever $\atrt$ is changed. As a by-product, we
automatically obtain samples of $\KK^{-1}\atr$ as required for\eq{eq:lgrad}.

Averages over the posterior distribution are taken by sampling after
each trajectory step, repeating the procedure over some large number
of steps.  In practice usually the first half of the steps are
discarded to allow for equilibration. We chose a total of 40,000
samples, giving 20,000 ``production samples'' with which to calculate
the averages needed for the calculation of the gradients,
eqs.~(\ref{eq:cgrad}) and~(\ref{eq:lgrad}).

\subsection{Gradient ascent algorithm}
\label{sec:gradient_ascent}

The numerical values for the gradient of the evidence, estimated as
explained above, were used in a simple gradient ascent algorithm to
move the hyperparameters to a local maximum of the evidence. More
powerful optimization techniques are not feasible in our case because
neither the evidence itself, nor the Hessian of the evidence are
available. The conjugate gradient method, for example, incorporates
information about the Hessian and also usually employs a line
search using the values of the function to be optimized.
Approximations to Newton's method such as the Levenberg-Marquardt
algorithm also use second derivative information.  Fortunately, using
the first derivatives of the evidence with respect to the
hyperparameters leads to convergence to some maximum in a reasonable
amount of time: typically between 40 and 80 steps of gradient ascent
are required before the gradients have shrunk to small values. 

For the experiments described the ``learning rate'' multiplier for the
derivative of each parameter is adapted separately throughout the
optimization.  This is necessary as the gradients vary over several
orders of magnitude during a typical simulation.  In our case the
adaptation of the``learning rate'' of the optimization must be based
on the change in the gradients only rather than on the change in the
evidence itself. We expect gradients to increase only at the start of
a simulation, but thereafter they should decrease as the parameters
approach a maximum in the evidence.  If the gradients do not decline
quickly then the learning rate is increased, if the gradients increase
sharply then the ascent step is discarded and the learning rate is
decreased.  For vector parameters (like the length scales in an RBF
kernel) the change in gradient diretion can also be used for learning
rate adaptation: sudden and large changes in the gradient suggest that
the optimization may have passed a maximum and the step should be
redone with a smaller learning rate. As in the experiments with zero
temperature Monte Carlo search, gradient ascent steps 
for the kernel hyperparameters were actually taken in the logarithms of these
parameters.

As noted above, the HMC simulation calculates averages over the
posterior with only a relatively small amount of noise.  Consequently,
for a given set of starting hyperparameters an optimization based on
gradient ascent in the evidence is practically deterministic.  So in
order to investigate the properties of local maxima in the evidence
repeated trials were performed with the SVM hyperparameters
initialized to random values. A few preliminary trials were used to
choose reasonable orders of magnitude, 
%
%
and unless specified otherwise
all results reported begin with uniform random initalization in the
ranges $C \in [0.4,0.8]$, $\ln l_a \in [-1,2]$ for all length
scales, $\ln k_0 \in [-1,1]$ and $\ln\koff \in [-2,-1]$.

\subsection{Computational effort}
\label{runtime}

We conclude this section with a brief discussion of the computational
demands of the various model selection methods; though we stress once
more that our focus was not on computational efficiency, so that
faster algorithms can almost certainly be designed for all of the
model selection criteria that we consider.

The computationally cheapest of the simple model selection criteria is
$\egacv$, which can be evaluated in time ${\mathcal O}(n)$ from the
properties of the trained SVM classifier. The span estimate $\espan$
requires the inversion of a matrix of size equal to the number of SVs;
assuming that the number of SVs is some finite fraction of $n$ for
large $n$ this gives a cost of ${\mathcal O}(n^3)$ for large $n$. The
Laplace approximations to the evidence, for both linear and quadratic
penalty SVMs, are dominated by the evaluation of determinants whose
size is also the number of SVs (or, for linear penalty SVMs, the
number of marginal SVs), giving again a scaling of approximately ${\mathcal
O}(n^3)$. 

Table~\ref{tab:time} lists the running times for a single
optimization step with each of the different methods. The evidence
approximations and the test error approximations showed more or less
similar running times, although the span estimate took somewhat longer
on average.  By far the greatest run time was needed for the gradient
ascent on the evidence, due to the HMC sampling involved; a typical
optimization run on a single processor HP V-Class took anywhere from 6
hours to 6 days.  In comparison, most optimizations based on the
simple model criteria were under an hour.  The run time of the HMC
algorithm should scale relatively benignly as $O(n^2)$ in the size of
the training set, but our experiments show that the prefactor is
large. The $n^2$ scaling comes mainly from the conversion from $\atrt$
to $\atr$ via $\atr=\KK\atrt$ which is necessary during the solution
of the Hamiltonian equations. (The length of the Hamiltonian
trajectory, \ie\ the time $\tau$, does not need to be increased with
$n$; the same is true for the number of samples required to obtain the
posterior averages to a given accuracy.)

Note that the theoretical dependence of running time on training set
size was not strictly followed in reality.  One reason for this is
that the average time per step presented includes time spent on
discarded steps in the zero temperature Monte Carlo search algorithms.
That is, the speed of the simple optimization techniques used here
depends on the complexity of the search space.

As stated above, we were interested in the evidence
gradient ascent algorithm mainly as a baseline for SVM model selection
based on probabilistic criteria. Computational efficiency could
however be increased in a number of ways; the Nystr\"om
method~\cite{WilSee01}, for example, could significantly reduce the
dimensionality (currently $n$) of the space over which the posterior
needs to be sampled using HMC.

\section{Numerical results}
\label{sec:results}

\subsection{Data sets} 

The model selection methods under consideration were applied to five
two-class classification problems that are common in the machine
learning literature.  Three of these are from real world problems: the
Pima Indian Diabetes data set, the Crabs data set and the Wisconsin
Diagnostic Breast Cancer (WDBC) data set.  The remaining two data
sets, Twonorm and Ringnorm, are synthetic. The dimensionality of the
inputs $x$ and the size of the training and test sets for each data
set are given in Table~\ref{tab:data}.  All benchmark data sets are
available through the UCI Machine Learning Repository
(http://www1.ics.uci.edu/$\sim$mlearn/MLRepository.html) and/or the 
DELVE archive (http://www.cs.toronto.edu/$\sim$delve/).  More
detailed descriptions are also available on the web. Inputs were
standardized so that across each complete data set all input
components had zero mean and unit variance.  For each data set the
training and test sets were held constant for all experiments.  The
first $n$ points in the data set were used for training and the
remaining points were used for testing.  The one exception is the
Crabs data set, where the 6th attribute (color) was not used for
classification and the remaining points were sampled to ensure an even
distribution of the unused color attribute in the training and test
sets. The number of training points, given in Table~\ref{tab:data}, was
the same as that used in previous research.

\begin{table}
\begin{center}
\begin{tabular}{|l||c|c|c|}
\hline
Data set & Inputs & Training set size & Test set size \\
\hline\hline
Crabs & 5 & 80 & 120 \\
\hline
Pima & 7 & 200 & 332  \\
\hline
WDBC & 30   & 300 & 269 \\
\hline
Twonorm & 20 & 300 & 7100 \\
\hline
Ringnorm & 20 & 300 & 7100 \\
\hline
\end{tabular}
\caption{Number of input dimensions, and sizes of training and test
sets for the data sets used in our experiments.} 
\label{tab:data}
\end{center}
\end{table}

\begin{table}
\begin{center}
\begin{tabular}{|l||c|c|c|c|c|c|}
\hline
Data Set  & SVM & LE1 &  LE2 & GACV   & Span & Evid grad \\ 
\hline
Crabs     & 0.81    & 3  &  5  & 4   & 6  &  137 \\
Pima      & 1.23    & 10   & 9  & 11  & 21  & 1805  \\
WDBC      & 2.1    & 19  & 26 & 39  & 64  & 9352 \\
Twonorm   & 2.5    & 35 & 26  & 27  & 82  & 7779 \\
Ringnorm  & 3.7    & 58 & 71  & 68  & 216 & 10665  \\
\hline
\end{tabular}
\end{center}
\caption{Average CPU time (seconds) per optimization step. 
Times are given for: training of the SVM classifier (SVM);
evaluation of the Laplace approximation to evidence for $p=1$ and
$p=2$ (LE1, LE2); evaluation of $\egacv$ and $\espan$ (GACV, Span);
and evaluation of the evidence gradients (Evid grad)
}
\label{tab:time}
\end{table}

\subsection{Model selection using simple criteria}

We discuss first the results obtained by optimizing the four simple
model selection criteria: the Laplace evidence (LE), eq.\eq{eq:lapnv},
and the GACV\eq{eq:gacv} for linear penalty SVMs, and the Laplace
evidence\eq{eq:laplace_qu} and span error estimate\eq{span} for
quadratic penalty SVMs. The experiments with gradient ascent on the
evidence, for linear penalty SVMs, are described separately in
Sec.~\ref{sec:results_gradient_ascent}. 

\begin{figure}[hp]
\vspace*{-1cm}
\includegraphics[angle=0, width=1.0\textwidth]{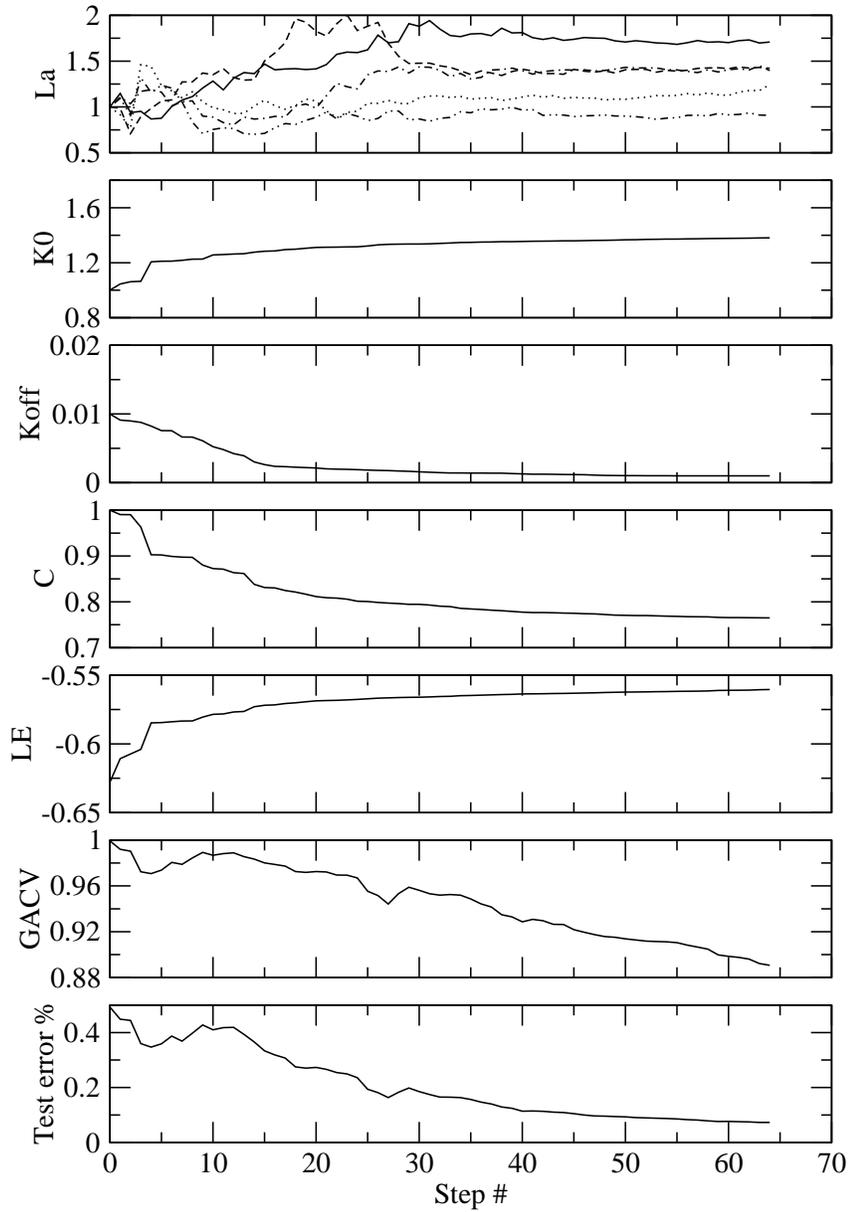}
\vspace*{-1cm}
\caption{Hyperparameter tuning for the Twonorm data set
by optimizing the Laplace approximation to the evidence for linear
penalty SVMs. The top four graphs show the evolution of the
hyperparameters; 5 out of the 20 length scale parameters are
shown. Below, the Laplace evidence (LE) is shown; the GACV error
estimate and the test error are also displayed, to demonstrate the
correlation with the Laplace evidence.
\label{fig:twonorm_lz}
}
\end{figure}

A typical example of selecting parameters for a linear penalty SVM by
optimizing the Laplace approximation of the evidence is shown in
Fig.~\ref{fig:twonorm_lz}.  The data set
for the experiment shown is the Twonorm data set.  This example is
chosen because it shows several typical features that appear in
similar forms in all of the optimizations. To what extent
optimizations for the other data sets match this example will be noted
where appropriate.

The parameters all move more or less stochastically to stable final
values as the evidence is optimized and it is clear that maximizing
the Laplace Evidence correlates to reducing the error on a test set.
Both the Laplace evidence and the GACV are shown alongside the test
error, although only the Laplace evidence is used for optimization.
Maximizing the evidence generally reduces the GACV, although
this correspondence is not strict. Similar behaviour is observed for
optimization with the Laplace evidence for the quadratic penalty SVM,
and for optimization of the GACV and the span estimate.  For quadratic
penalty SVMs, the Laplace evidence and the span estimate also have the
same qualitative correlation as the Laplace evidence and the GACV for
the linear penalty case.

An important issue in all of these methods is the existence of many
local optima in the model selection criteria. Starting from the same
initialization, the hyperparameters converged to significantly
different values in repeated trials. We verified explicitly, \eg\ by
evaluating the chosen model selection criterion along a line in
hyperparameter space connecting different end points of two
optimization runs, that the different local optima found were genuine
and not artefacts due to incomplete convergence of the optimization
algorithms.  The search criteria always deteriorated in
between the points found by the search, confirming
that the latter were in fact local optima.

To analyse the characteristics of the
local optima, 25 repeated trials were performed on all data sets for the simple
optimization criteria.  Comparison of the final SVM hyperparameter
values at the local optima showed that they were highly variable.  For
all methods and all data sets the variance of the final parameter
values was always of the same order of magnitude as the average value
of the final parameters.  Tuning of the length scales is often interpreted
as ``relevance determination'' for the different dimensions of the
data because a large length scale indicates that the classification
does not vary significantly with changes in that parameter.  The
results here however indicate that the relevance of each dimension
probably depends in a complicated way on the relevance assigned to the
other dimensions, and that different assignments of the length
scales can yield similar results.

In addition to variance in the final SVM hyperparameter values, the
test error also showed significant trial to trial variation.  For all
methods, many of the trials result in a final test error that is close
to the best achieved by any method, but for some methods a large
portion of the trials end in a test error that is significantly worse.
The average and standard deviation of the test errors achieved with
the different methods are shown in Table~\ref{tab:results}.
Table~\ref{tab:bestresults} shows the best test errors achieved on any
trial for each method and data set.  For reference,
Table~\ref{tab:prevresults} shows the test errors achieved on the
same data sets by comparable methods in previous research.
Unfortunately, these previous studies do not always include error bars for
test error results so it is hard to compare the results for the
averages and standard deviations of the error.

\begin{figure}[hp]
\includegraphics[angle=0, width=1.0\textwidth]{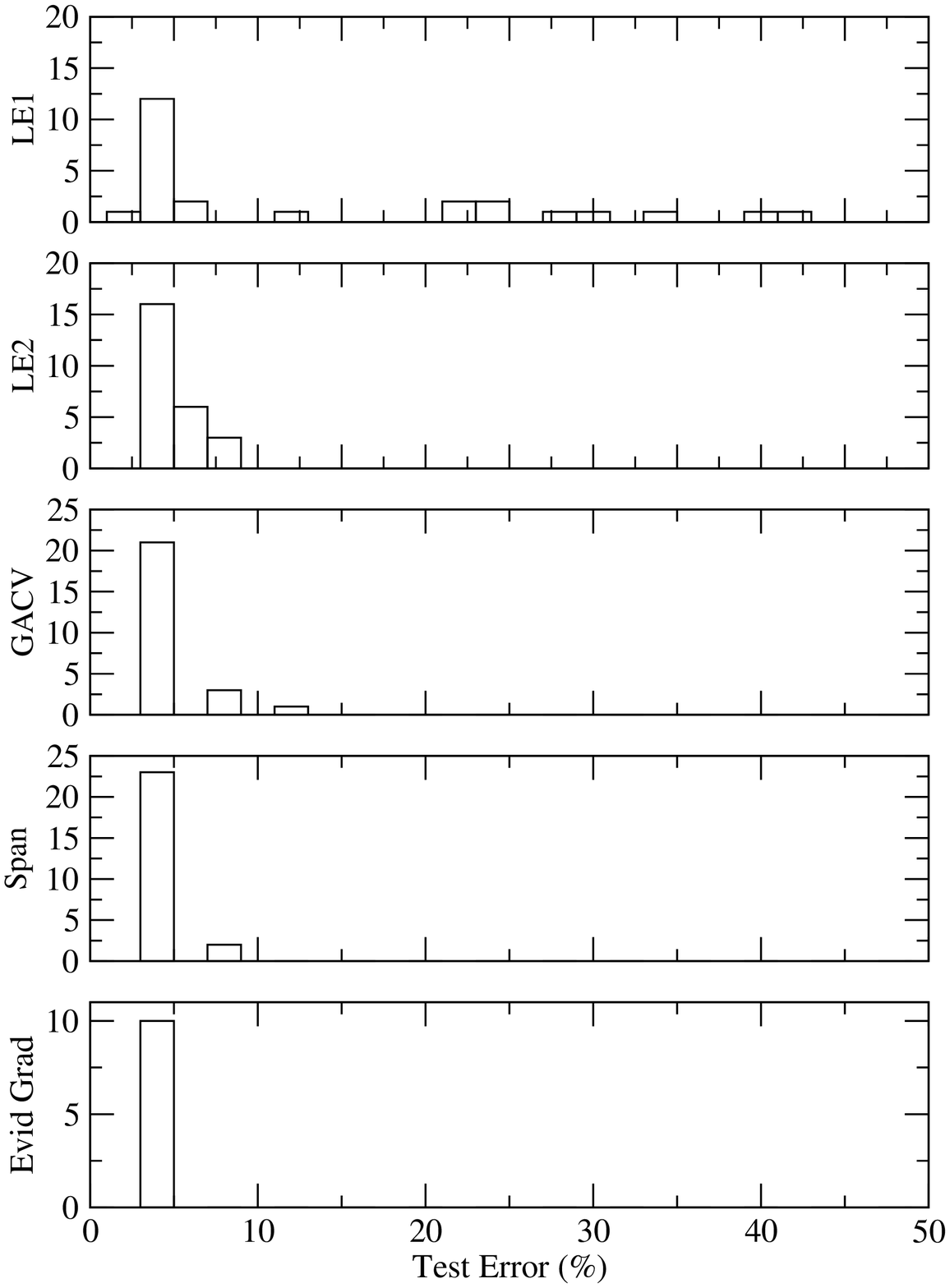}
\vspace*{-1cm}
\caption{Histogram of test errors (in $\%$) achieved on the Twonorm
data set. Shown are the results of 25 trials with the simple model
selection criteria, and 10 trials of evidence gradient ascent.  The bin
size for the histogram is 2\%. }
\label{fig:twonorm_eh}
\end{figure}

\begin{figure}[hp]
\includegraphics[angle=0, width=1.0\textwidth]{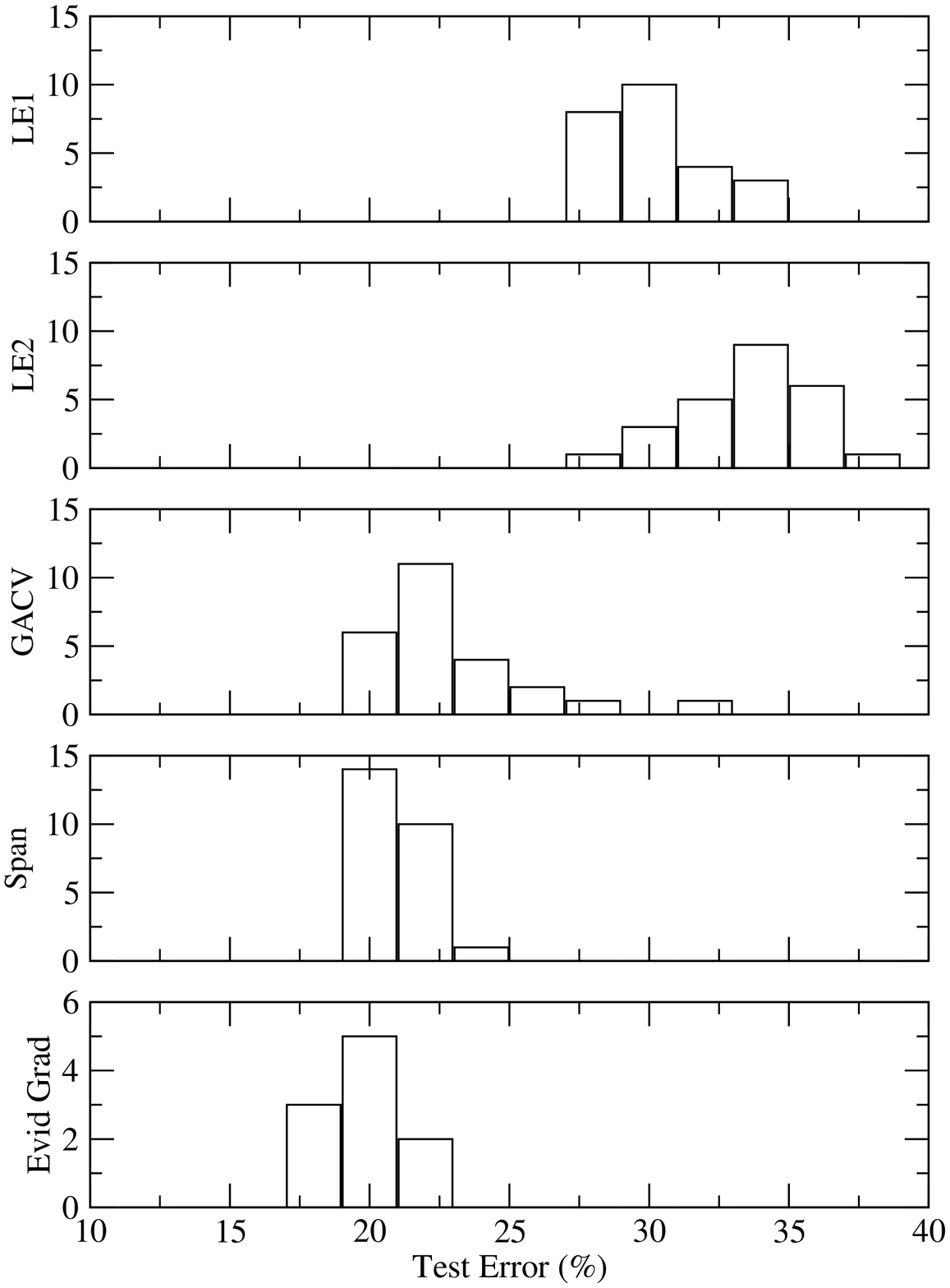}
\vspace*{-1cm}
\caption{Histogram of test errors (in $\%$) achieved on the Pima
data set. Shown are the results of 25 trials with the simple model
selection criteria, and 10 trials of evidence gradient ascent.  The bin
size for the histogram is 2\%.  }
\label{fig:pima_eh}
\end{figure}

\begin{figure}[hp]
\includegraphics[angle=0, width=1.0\textwidth]{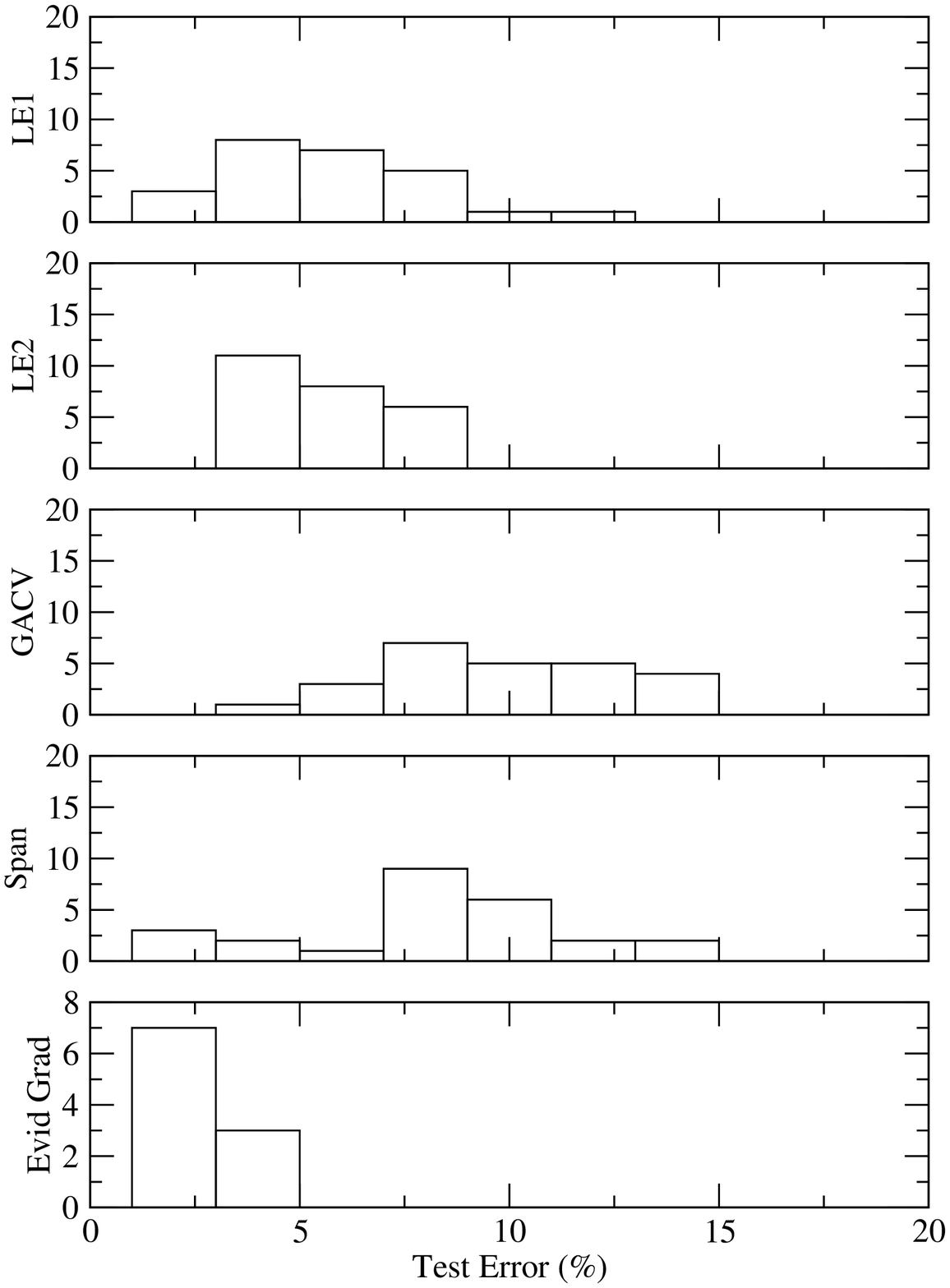}
\vspace*{-1cm}
\caption{Histogram of test errors (in $\%$) achieved on the WDBC
data set. Shown are the results of 25 trials with the simple model
selection criteria, and 10 trials of evidence gradient ascent.  The bin
size for the histogram is 2\%.}
\label{fig:wdbc_eh}
\end{figure}

To illustrate the variability in the final error resulting from each
optimization method histograms of the errors achieved in all trials
are shown for the Twonorm, Pima and WDBC data sets in
Figs.~\ref{fig:twonorm_eh},~\ref{fig:pima_eh}, and~\ref{fig:wdbc_eh}
respectively.  These plots show the difficulty of picking a ``best''
method from among the simple model selection criteria.  For the
Twonorm data set all of the methods produce test errors that are
around the best for any method in previous research, but with the
evidence approximation for linear penalty SVMs around a third of the trials
end in errors
that are signficantly greater.  For the Pima data set, all of the
simple methods are inferior to the best methods in previous research,
and both of the evidence approximations perform worse than the error
estimates; while on the WDBC data set the evidence approximations are
superior to the error estimates.

Comparing Tables ~\ref{tab:results}, ~\ref{tab:bestresults},
and~\ref{tab:prevresults}, it is clear that the high variability in
the results achieved by optimizing the four ``simple'' model selection
criteria is undesirable; while the best trials for each method and
data set are approximately the same as the best results reported in
previous research, the average performance over trials is rather
disappointing. One possible productive use of the high variability of
classifiers produced by convergence to local optima of the model
selection criteria could be to combine the resulting classifiers in
some ensemble or voting scheme. Such approaches normally benefit
precisely from high variability among the classifiers being combined,
so this could be an interesting subject for future research.

\begin{table}
\begin{center}
\newcommand{\xx}{\hspace{1mm}}
\begin{tabular}{|l||r@{\xx}c@{\xx}r|r@{\xx}c@{\xx}r|r@{\xx}c@{\xx}r|r@{\xx}c@{\xx}r|r@{\xx}c@{\xx}r|r@{\xx}c@{\xx}r|}
\hline
&
\multicolumn{3}{c|}
{LE1} &
\multicolumn{3}{c|}
{LE2} &
\multicolumn{3}{c|}
{GACV} &
\multicolumn{3}{c|}
{Span} &
\multicolumn{3}{c|}
{Evid grad}&
\multicolumn{3}{c|}
{ES} \\ 
\hline\hline
Crabs  &  10.7  &$\pm$&  2.1  &  10.5  &$\pm$&  1.3  &  13.0  &$\pm$&  1.8  &  6.0  &$\pm$&  1.8  &  9.2  &$\pm$&  1.5 &  5.5  &$\pm$&  1.3  \\ 
\hline
Pima  &  30.3  &$\pm$&  2.0  &  33.5  &$\pm$&  2.2  &  23.2  &$\pm$&  2.7  &  21.0  &$\pm$&  1.1  &  20.8  &$\pm$&  1.5 &  19.7  &$\pm$&  1.5 \\
\hline
WDBC     &  5.8  &$\pm$&  2.5  &  5.8  &$\pm$&  3.6  &  9.6  &$\pm$&  2.8  &  7.8  &$\pm$&  4.2  &  4.0  &$\pm$&  1.2 &  2.4  &$\pm$&  1.0 \\
\hline
Twonorm  &  13.5  &$\pm$&  12.6  &  12.6  &$\pm$&  5.0  &  5.2  &$\pm$&  1.9  &  4.6  &$\pm$&  0.9  &  4.0  &$\pm$&  0.2 & 3.7  &$\pm$& 0.4  \\
\hline
Ringnorm  &  4.7  &$\pm$&  1.6  &  2.5  &$\pm$&  5.3  &  3.3  &$\pm$&  1.3  &  3.5  &$\pm$&  1.3  &  3.2  &$\pm$&  0.6 &  3.2  &$\pm$&  0.6 \\ 
\hline
\end{tabular}
\end{center}

\caption{Test error $\epsilon$ for all data sets (in $\%$), written in
the form ``mean $\pm$ standard deviation''.  Statistics for the simple
model selction criteria are taken over 25 trials.  For gradient ascent
in the evidence averages are over 25 trials for the Crabs data set,
and over 10 trials for all other data sets. Abbreviations for the
model selection criteria are as in Table~\protect\ref{tab:time},
except for the last column (ES = gradient ascent with ``early
stopping''; see Sec.~\protect\ref{overfitting}).
\label{tab:results}
}
\end{table}

\begin{table}[h]
\begin{center}
\begin{tabular}{|l||c|c|c|c|c|c|}
\hline
& LE1 & LE2 & GACV & Span & Evid grad & ES \\ 
\hline\hline
Crabs & 5.9 & 9.2 & 10.9 & 3.4 & 5.0 & 3.4 \\
\hline
Pima & 27.8 & 28.4 & 20.2 & 19.0 & 19.3 & 18.4 \\
\hline
WDBC & 1.9 & 3.4 & 4.1 & 1.9 & 1.5 & 1.2 \\ 
\hline
Ringnorm & 2.1 & 2.2 & 1.9 & 2.0 & 2.5 & 2.5 \\
\hline
Twonorm & 2.9 & 3.1 & 3.4 & 3.5 & 3.4 & 3.0 \\
\hline
\end{tabular}
\end{center}
\caption{Best single trial test error $\epsilon$ (in $\%$).
Abbreviations for the model selection criteria are as in
Table~\protect\ref{tab:results}.
\label{tab:bestresults}
}
\end{table}

\begin{table}
\begin{center}
\begin{tabular}{|l||c|c|c|c|c|}
\hline
Data set & GP Var & SVM Var & SVM CV & GP Lap & GP MF \\
\hline\hline
Crabs    & 2.5    & 3.3     &  3.3   & 3.3   & 1.7 \\
\hline
Pima     & 19.9   & 20.5    &  20.2  & 20.2  & 19.0 \\
\hline
WDBC     & 3.7    & 3.7     &  3.3   & 3.3   & 2.6 \\
\hline
Twonorm  & 3.2    & 3.7     &  2.3   & 4.0   & --- \\
\hline
Ringnorm & 1.7    & 1.9     &  2.3   & 3.0   & --- \\
\hline
\end{tabular}
\end{center}
\caption{Test errors $\epsilon$ (in $\%$) found on the benchmark data
sets in previous work. The methods used were as follows.
GP Var: Gaussian process classifier (see \eg~\protect\cite{BarWil97,WilBar98}),
with hyperparameters determined by maximizing a variational
approximation to the evidence~\protect\cite{Seeger00}.
SVM Var: SVM with hyperparameters selected by the same variational
method~\protect\cite{Seeger00}.
SVM CV: SVM, with all length scales $l_a=l$ set equal and $l$ and
$k_0$ determined by ten-fold cross validation~\protect\cite{Seeger00};
the offset was unrestricted, \ie\ effectively $\koff\to\infty$.
GP Lap: Gaussian process classifier, hyperparameters determined by
maximizing a Laplace approximation to the
evidence~\protect\cite{Seeger00};
GP MF: Gaussian process classifier trained by a mean
field method~\protect\cite{OppWin00}.
}
\label{tab:prevresults}
\end{table}

\subsection{Model selection using evidence gradients}
\label{sec:results_gradient_ascent}

\begin{figure}[hp]
\begin{center}
\includegraphics[width=\textwidth]{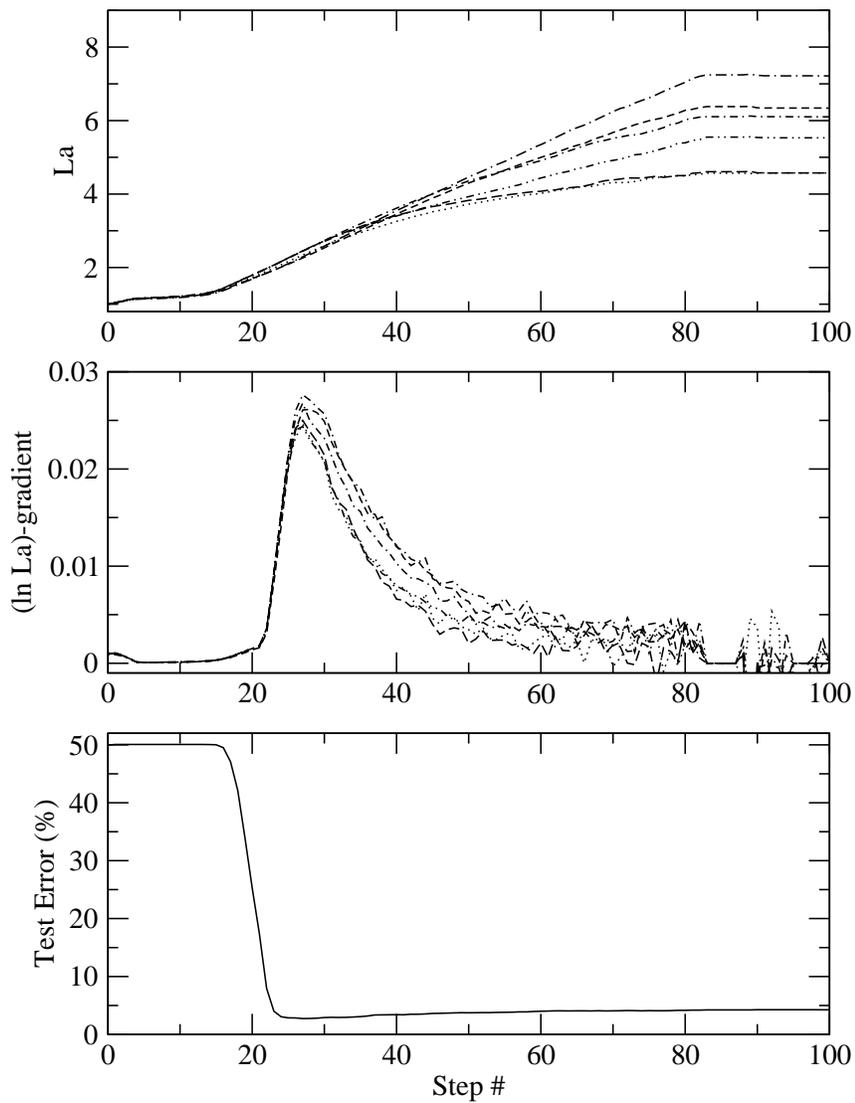}
\end{center}
\vspace*{-1cm}
\caption{Tuning the length scales $l_a$ on the Twonorm data set, using
gradient ascent on the evidence. 6 out of 20 length scale parameters
are shown, along with the corresponding gradients; the bottom plot
shows the evolution of the test error. 
\label{fig:twonorm_ll}
}
\end{figure}

\begin{figure}[hp]
\hspace*{-0.15\textwidth}
\includegraphics[angle=270, width=1.35\textwidth]{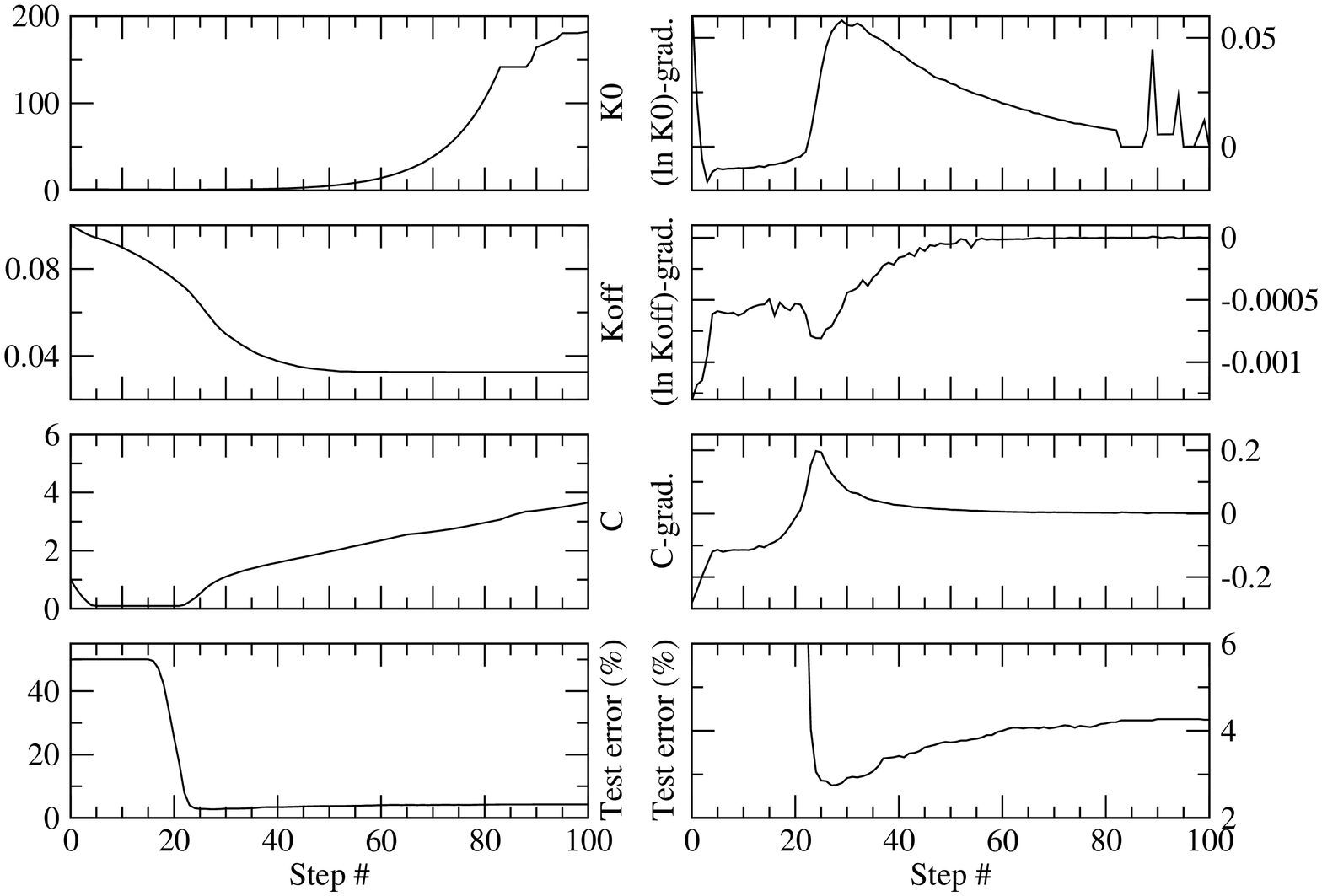}
\caption{Tuning $k_{0}$, $\koff$ and $C$ on the Twonorm data set,
using gradient ascent on the evidence.   The gradients for each
parameter are shown alongside the actual parameter values.  The
two bottom panels show the evolution of the test error, with the right
one being a zoom on the range of small error values; see discussion
in Sec.~\ref{overfitting}.
}
\label{fig:twonorm_kkc}
\end{figure}

Figs.~\ref{fig:twonorm_ll} and~\ref{fig:twonorm_kkc} show a typical
run of evidence gradient ascent on the Twonorm data set.
Fig.~\ref{fig:twonorm_ll} shows the tuning of a subset of the RBF
kernel length scales and Fig.~\ref{fig:twonorm_kkc} shows the tuning
of kernel amplitude $k_0$, the kernel offset $\koff$ and the penalty
parameter $C$.  (Although statistics for the performance of the evidence
gradient method were determined by initialization to random parameter
values, for the specific sample shown we started all length scales with
identical parameters.) Both the gradients of the evidence with
respect to each parameter and the parameter values themselves are
shown.  The gradients typically start at small values, rise to a peak
and then decline. Most parameter ultimately arrive at a constant value
with small gradients, indicating that the evidence is at a local
maximum with respect to that parameter. The optimization is terminated
when the gradients have reached a small fraction of their peak
magnitude.  During this process the error on the test set decreases
significantly.

As with the simple model selection criteria analysed in the previous
section (Laplace approximations to the evidence and error
approximations), repeated trials of gradient ascent in the evidence
showed the existence of many local maxima in the evidence at widely
varying parameter values.  Due to the long run time required for the
HMC sampling used to calculate the evidence gradients only 10 trials
were performed for each benchmark data set, with the exception of the
Crabs data set where 25 trials were performed.  (See Section
\ref{runtime} for a discussion of the running time of the algorithm on
the various data sets.) As explained in
Sec.~\ref{sec:gradient_ascent}, the gradient ascent algorithm is
essentially deterministic once the initial hyperparameter values have
been fixed.  Consequently, repeated trials were started from random
initial values for the hyperparameters in order to investigate the
existence and variability of local maxima in the evidence.

Tables~\ref{tab:results} and~\ref{tab:bestresults} above list the
resulting test errors obtained with gradient ascent optimization of the
evidence, along with results obtained from the simpler methods discussed
earlier. Comparing with the results found in previous studies (Table
~\ref{tab:prevresults}), one sees that gradient ascent on the evidence
for SVMs with radial basis function kernels achieves approximately the
same test error as the best methods that have been previously applied.
For some data sets the best performance obtained by evidence gradient
ascent is superior to the performance previously reported.  An
interesting point of comparison is with SVM model selection by
optimization of a variational approximation of the evidence, as
described in~\cite{Seeger00}. (This comparison is somewhat tentative
because of the small number of trials for our evidence gradient ascent
method, combined with the lack of information about trial-to-trial
variation in~\cite{Seeger00}.) Still, it is worth noting that
although the method desribed here uses gradients of the evidence
without further approximation, and also tunes the $C$ parameter (which
is effectively fixed to unity in the approach of~\cite{Seeger00}), it
does not seem to achieve systematically superior performance than the
variational approximation.

Figs.~\ref{fig:twonorm_eh}, \ref{fig:pima_eh} and~\ref{fig:wdbc_eh}
above contain the histograms of the test error produced by SVM model
selection by evidence gradient ascent, and the comparison with the
simple model selection criteria, for the Twonorm, Pima and WDBC data
sets.  Although strong conclusions cannot be drawn due to the small
number of trials, gradient ascent in the evidence seems to produce
significantly better performance on all of the data sets than any of
the other methods.  In all tests the distribution of resulting errors
is both closer to the best results found in previous studies, and less
variable.  The most likely explanation for the superior performance of
the gradient ascent method is the fact that it actually maximizes
an exact, unapproximated model selection criterion (the evidence),
while the simple model selection criteria (Laplace evidence
and error estimates) are all to some extent approximate. The poorly
performing local optima of these simple criteria may then arise from
errors introduced by the approximations.

We comment briefly on the actual values of the hyperparameters found
by gradient ascent on the evidence, in particular for the kernel
amplitude $k_0$ and the offset $\koff$. For the Twonorm data set, the
maximum in the evidence occurs at a relatively large value of $k_0$,
with an average of $k_0\approx 180$ across trials. (Large values of
$k_0$ were also found with model selection with the approximate
evidence, but were much less common when using the error estimates.)  This may
appear surprising. However, it should be born in mind that from the
probabilistic view the prior variance of the latent function $\theta$
is $\lav\theta^2(x)\rav=K(x,x)=k_0+\koff$. The typical prior scale for
$\theta(x)$ is therefore $\sqrt{k_0}$ (since $\koff$ is small, see
below), which equates to around 13 for $k_0\approx 180$; this is not
unreasonably large compared to the scale of 1 set by the SVM margin.
Similar final values of $k_0$ were obtained for the Crabs and WDBC
data sets, while for Pima and Ringnorm $k_0$ was rather smaller.
Previous experiments with simple synthetic data sets~\cite{Sollich02}
suggest that an evidence maximum at large $k_0$ correlates with small
apparent levels of noise in the data set; we have not attempted to verify this
correlation for our five benchmark data sets.

The offset hyperparameter $\koff$ was typically tuned to very small
values by evidence gradient ascent (\eg\ around 0.03 for the Twonorm
data set). This provides a posteriori justification for our approach
of including the offset parameter $b$ from the conventional SVM
framework into the kernel.

\subsubsection{Noise in evidence gradients}

In the final portions of the optimization shown in
Figs.~\ref{fig:twonorm_ll} and~\ref{fig:twonorm_kkc} it can be
observed that there is significant noise in the gradients as the
evidence approaches a maximum.  This arises from statistical
fluctuations in the HMC sampling, which come to dominate when the true
gradient values are small. Although the noise could be decreased by
increasing the length of the HMC runs, that did not seem to be
necessary for the cases considered here: because of the learning rate
adaptation the learning rate is quite small by the time the evidence
is close to its maximum and the noise in the gradients has little
effect on the final results.

In the Twonorm example it can also be seen that when the parameters
are nearly at a maximum in the evidence the gradients with respect to
the kernel parameters are calculated as zero in some steps.  This
effect occurred typically for larger values of $C$. Regions of
$\atr$-space where the potential $V(\atr)$ in the
Hamiltonian\eq{eq:ham} is zero, \ie\ where all $\y\theta_i\geq 1$, are
then much more probable then regions where $\y\theta_i<1$ for some
$i$. It is then possible that the HMC sampling only returns samples
from the region with $V(\atr)=0$, where $l'_p(\y\theta_i)=0$ for all $i$ so
that\eq{eq:lgrad} gives an estimate of zero for all gradients with
respect to kernel parameters.  Experiments showed that scaling the
trajectory length in the HMC runs proportionally to $\frac{1}{C}$ for
large $C$ could avoid this effect. The rationale is that
the shorter trajectories makes the HMC sampling more likely to sample
values of $\atr$ 
which are just outside the boundary of the $V(\atr)=0$ region; these
still have appreciable posterior probability but do give nonzero
values for some of the $l'_p(\y\theta_i)$.  We did not explore this
issue in detail, however.

\subsubsection{``Overfitting'' by evidence maximization}
\label{overfitting}

Close inspection of progress of the test error in
Figs.~\ref{fig:twonorm_ll} and~\ref{fig:twonorm_kkc} shows an
interesting aspect of tuning SVM hyperparameters using the evidence.
While the overall evolution of the test error shows a large decline as
gradient ascent on the evidence progresses, a closer look at the
region of small error values (see the lower right plot of
Fig.~\ref{fig:twonorm_kkc}) shows that the test error goes through a
shallow minimum before a small rise to its final value.  Not all data
sets show such a clean example of this behaviour as the Twonorm
data set, but all except Pima did exhibit the phenomenon to some
degree.

One possible explanation for the observed test error minimum is the
fact that we are not using 
the evidence of a properly normalized probabilty model (see
Sec.~\ref{sec:svm}). An alternative interpretation, which seems to us
more likely, is that we are observing here a kind of overfitting. This
takes place not on the level of the ``network'' parameters ($\w$ or
$\theta(x)$) as in conventional overfitting -- which is due to a lack of
regularization -- but on the level of the hyperparameters: If we imagine
sampling a number of data sets of size $n$ from a given true
distribution, then the evidence as a function of the hyperparameters,
and hence the position of its maximum, will depend on the particular data
set. Only for large $n$ would the evidence become independent of the
data set (and related to the Kullback-Liebler divergence, or
cross-entropy, between the true distribution over data sets and the
one predicted by the inference model; see \eg~\cite{Sollich02}). For
finite $n$, maximization of the 
evidence for a specific data set is therefore not expected to lead to
strict minimization of the error on an independent test set.

This interpretation leads naturally to the idea of using an early
stopping mechanism when optimizing the evidence, where the gradient
ascent is abandoned when performance on an independent validation set
ceases to improve. Note that this is not the same as simply returning
to hyperparameter tuning by cross-validation; in fact, a grid search
using cross-validation error over the large number of hyperparameters
in our examples ($C$, $k_0$, $\koff$ and the length scales $l_a$
associated with each of the $d$ input dimensions) would be essentially
impossible (see also~\cite{ChaVapBouMuk02}). To gauge the possible
benefits of such an approach, we have included in
Table~\ref{tab:results} above both the final test error when the
optimization is run until the gradients are small, and the minimal
value of the test error during the gradient ascent. True early
stopping with an independent validation set would be expected to yield
a performance in between these two values; the results in
Table~\ref{tab:results} suggest that this could be useful for some
data sets.

\section{Conclusion}
\label{sec:conclusion}

In this paper we have investigated the issue of model selection for
SVM classifiers. We have restricted ourselves to model selection in
the sense of tuning the parameters of an RBF kernel and the penalty
parameter $C$, though the general approaches described could also be
used for choosing between different functional forms of the kernel.

We reviewed briefly the probabilistic view of SVMs, and extended our
previous work on Laplace approximations to the evidence to the case of
SVMs with quadratic slack penalties. Exact expressions for the
gradients of the evidence in terms of posterior averages were also
derived, and we described how these averages can be estimated
numerically using Hybrid Monte Carlo techniques and used in a model
selection algorithm which performs gradient ascent on the exact
(unapproximated) evidence.

In our numerical experiments on five benchmark data sets, we compared
optimization of four ``simple'' model criteria with the evidence
gradient descent. Two of the simple criteria were estimates of test
error: the generalized approximate cross-validation error (GACV) for
SVMs with linear slack penalties, and the span error estimate for SVMs
with quadratic penalties. The two other criteria were derived from
probabilistic concepts; these were the Laplace approximations to the
evidence for the linear and quadratic penalty cases. Our main result
is that all the simple model criteria exhibit multiple local optima
with respect to the hyperparameters. While some of the resulting
``locally optimal'' SVM classifiers give test performance that is
competitive with results from other approaches in the literature, a
significant fraction lead to rather higher test errors. The results
for the evidence gradient ascent method show that also the exact
evidence exhibits local optima. But these give much less variable test
errors, which are also typically lower test errors than for the
simpler model selection criteria. In this sense, ``you get what you
pay for'': the computationally rather more expensive evidence gradient
ascent approach gives better and more consistent performance than the
cheaper model selection criteria.

There are a number of directions for possible future work. First, our
results strongly suggest that the hunt is still on for a model
selection criteria for SVM classification which is both simple and
gives consistent generalization performance. Alternatively,
one could try to cope with the existence of local maxima in the simple
model selection criteria by testing the selected models on a
validation set and performing repeated optimizations until satisfactory
performance is found. A more interesting approach might be to try to
exploit the large variability in the locally optimal classifiers by
using them in some scheme for combining classifiers. Finally, if
evidence gradient ascent turned out in more comprehensive tests to be
the model selection method of choice, it would be worth investigating
possible speed-ups of the algorithm. We already hinted at the Nystr\"om
method~\cite{WilSee01} above, but one could also explore running the
model selection only on randomly sampled subsets of data, and then
possibly combining the resulting classifiers appropriately.

\subsubsection*{Acknowledgements}

Access to the Hewlett-Packard V2500 was provided by the Caltech Center
for Advanced Computing Research (http://www.cacr.caltech.edu) through
the National Partnership for Advanced Computational Infrastructure--A
Distributed Laboratory for Computational Science and Engineering,
supported by the NSF cooperative agreement ACI-9619020.  We also thank
Andrew Buchan for assistance with the early stages of the numerical
experiments for quadratic penalty SVMs.

\bibliographystyle{unsrt}
\bibliography{references}

\end{document}